\documentclass[preprint,aps]{revtex4}
\usepackage{graphicx}
\usepackage{epstopdf}
\usepackage{dcolumn}
\usepackage{bm}
\usepackage{color}

\begin{document}

\title{Specific many-electron effects in X-ray spectra of simple metals and graphene}

\author{R.\,E.~Ovcharenko,$^{1,}$\footnote{Corresponding author. E-mail: r.e.ovcharenko@gmail.com} I.\,I.~Tupitsyn,$^1$ E.\,P.~Savinov,$^1$ E.\,N.~Voloshina,$^2$ B.~Paulus,$^2$ Yu.\,S.~Dedkov,$^3$ A.\,S.~Shulakov$^1$}

\affiliation{$^1$V. A. Fock Institute of Physics, Saint-Petersburg State University, Ulianovskaya 1, 198504 Saint-Petersburg, Russia}
\affiliation{$^2$Institut f\"ur Chemie und Biochemie - Physikalische und Theoretische Chemie, Freie Universit\"at Berlin, Takustra\ss e 3, 14195 Berlin, Germany}
\affiliation{$^3$SPECS Surface Nano Analysis GmbH, Voltastra\ss e 5, 13355 Berlin, Germany}

\date{\today}

\begin{abstract}
In this work the influence of many-electron effects on the shape of characteristic X-ray emission bands of the simple metals Mg and Al are examined by means of \textit{ab~initio} calculations and semi-empirical models. These approaches are also used for the analysis of C K-emission and absorption spectra of graphene. Both, the dynamical screening of the core vacancy and the Auger-effect in the valence band (VB) have been taken into account. Dynamical screening of the core vacancy by valence electrons (the so-called MND effect) is considered \textit{ab~initio} in the framework of density functional theory. The Auger effect in VB was taken into account within a semi-empirical method, approximating the quadratic dependence of the VB hole level width on the difference between the level energy and the Fermi energy. All theoretical spectra are in very good agreement with available experimental data.
\end{abstract}

\maketitle

\section{Introduction}

The core-level spectroscopy techniques, x-ray emission and absorption spectroscopy (XES and XAS), are two complementary methods, allowing to get the full information about the electronic structure of the system (solids, molecules in gas or adsorbed phase, surfaces, interfaces, etc.) in a whole range of occupied/unoccupied valence states, including the vicinity of the Fermi level, i.\,e. the energy distributions of occupied and unoccupied electronic states, respectively~\cite{Groot:2001,Groot:2008}. Recently these methods became especially interesting and attractive with the development of the next generation of x-ray sources, the x-ray free-electron lasers~\cite{Alonso-Mori:2012}, and the further development of the spectroscopic techniques allowing to study collective excitations in solids~\cite{Ament:2011}. Spectral photon intensity distribution in characteristic X-ray emission bands (XEB) contains information on the local characteristics of the valence band (VB) electron density of states (DOS) distribution in crystals. Despite a considerable success of one-electron methods in the description of the electronic structure of metals, for reliable interpretation of features of the experimental XEB it is necessary to use  theoretical methods allowing, along with one-electron description of the XEB shapes, to consider many-electron effects which are consequences of the dynamical processes accompanying the X-ray transition. In particular, L$_{2,3}$-XEBs of simple Na, Mg and Al metals contain features which are poorly described by the one-electron model and are attributed to the many-electron processes. 

For the first time, the theoretical consideration of the XEB features of metals near the Fermi edge, caused by many-electron processes, was undertaken in Refs.~\cite{Mahan_67,Anderson_67}. It was shown that in the characteristic L$_{2,3}$-XEB and L$_{2,3}$-absorption spectra of simple metals a narrow peak must be present in a small energy range near the Fermi energy ($E_\mathrm{f}$), while in K-XEB, this feature should be absent. In Ref.~\cite{Mahan_67} this effect was associated with a dynamical screening of the core vacancy during the X-ray radiative transition. This vacancy occurs in the initial state of the emission transition or in the final state in the process of absorption. The model proposed in Ref.~\cite{Mahan_67} was later generalized in Ref.~\cite{Nozieres_69} and is presently known as the MND (Mahan-Nozieres-De Dominicis) theory. It qualitatively describes the behaviour of the X-ray emission and absorption spectra near $E_\mathrm{f}$, in the range of about $1$\,eV. In Refs.~\cite{Mahan_67,Anderson_67, Nozieres_69} a number of semi-empirical parameters were introduced, which were chosen to provide the best fit of the theoretical spectrum to the experimental one. In Refs.~\cite{Grebennikov_77a, Grebennikov_77b} a quantitative one-band scheme of the MND calculation was proposed. Only recently~\cite{Wessely_05, Wessely_06}, the MND theory was successfully applied for the calculation of the X-ray absorption spectra of graphite, where the electronic structure obtained from density functional theory (DFT) calculations was used as an input. 

Another feature of the XEB of metals, governed by many-electron processes, is a prolonged ``tail'' in the low-energy part of the emission spectrum~\cite{Landsberg_49, Nemoshkalenko_74}. It is accepted that this feature is a consequence of the nonradiative Auger process in the valence band of the metal. Indeed, the vacancy, appeared in the VB after the X-ray emission transition, can decay due to the Auger process characterising by the excitation of one more VB electron into conduction band above the Fermi level. When moving from the Fermi level to the bottom of the VB a lifetime of the hole in the final state of the emission transition decreases and the energy width increases. To date non-empirical \textit{ab~initio} treatment of this effect is too complicated and time consuming. Therefore in this study we used a semi-empirical method~\cite{Livins_88}, which allows to reproduce accurately the shape of metal XEB in low-energy range of a spectrum.

According to the original MND theory, in the free-electron approximation the dynamical screening of a core hole by valence electrons has to affect the shape of L$_{2,3}$ XEB, but should not significantly distort the K-absorption and emission spectra~\cite{Mizuno_68, Ohtaka_90}. In the case of K-spectra MND singularity near the Fermi level is suppressed because of the orthogonality between the VB many-electron wave function before and after an X-ray transition (Anderson's theorem~\cite{Anderson_67}). At the same time, in Refs.~\cite{Wessely_05, Wessely_06} it was shown that the MND effect contributes significantly to the spectral distribution of the absorption coefficient of graphite near the K-threshold of carbon. However, the experimental K-XEB of  carbon in graphite~\cite{Muller_73} does not contain signatures of the MND singularities. Therefore we calculated the K  XEB of graphene to determine the degree of influence of the $1s$ core hole dynamical screening on the shape of this band.

In this paper we present the results of \textit{ab~initio} calculations of XEB accounting for the MND scattering employing the computational scheme presented in Refs.~\cite{Grebennikov_77a, Grebennikov_77b, Wessely_05, Wessely_06}. Here \textit{ab \mbox{initio}} calculations of the K- and L$_{2, 3}$- XEB of the simple metals, as well as K- XEB of carbon in graphene were carried out  including the MND effect for the first time. The calculated XEB were compared with the experimental results.

\section{\label{par2} The shape of X-ray emission band with account of MND scattering}

Let us consider the radiative X-ray transitions of an electron from the band states $n \mathbf{k}$ on the core vacancy, localized on the atom $A$ with the quantum numbers $c \equiv \{n_c, l_c, m_c\}$. In the tight-binding approximation, neglecting the ``cross'' transition contribution to the intensity of radiation, the XEB intensity $I(E)$ can be represented as the sum of the products of the electric dipole probabilities of transition $W_{\mu}$ from the atom-like valence band state $\mu$ on the core vacancy and partial density of states $N_{\mu}(E)$
\begin{equation}
I(E) \sim \sum_{\mu} W_{\mu} \, N_{\mu}(E) \,,
\end{equation}
\begin{equation}
W_{\mu} \sim \sum_{m_c, \alpha} |A^{\alpha}_{\mu m_c}|^2 \,,
\label{probab1}
\end{equation}
where the amplitude $A^{\alpha}_{\mu m_c}$ of transition is given by
\begin{equation}
A^{\alpha}_{\mu m_c}=\langle \varphi_{\mu}|r_{\alpha} |\varphi_c \rangle \,.
\end{equation}
Here $\varphi_{\mu}$ and $\varphi_c$ are the atomic wave functions of  valence electron and core hole, respectively. In the framework of the one-particle approximation, the partial density of states (PDOS) can be expressed via the imaginary part of the advanced Green's function projected onto the atomic orbitals $G^{-}(E)$
\begin{equation}
N_{\mu}(E) = \frac{1}{\pi} \, \mathrm{Im} G^{-}_{\mu \mu}(E)\,.
\label{dens1}
\end{equation}
%
Completely ignoring the influence of the core hole, the Green's function can be written as
\begin{equation}
\label{G_final}
G^{-}_{\mu \nu}(E) = \sum_{n, \mathbf{k}} 
\frac{\langle \mu | \psi_{n \mathbf{k}} \rangle \,
\langle \psi_{n \mathbf{k}} | \nu \rangle}
{E-E_n(\mathbf{k})-E_{\rm f}-i \delta}\,,
\end{equation}
with $E_{\rm f}$ -- the Fermi energy, $E_n(\mathbf{k})$ and $\psi_{n \mathbf{k}}$ -- the unperturbed band energy and wave function, respectively. 
It may be considered as an assumption that the lifetime of the core hole is much smaller than the relaxation time of the system (an approximation of the final state).

Another extreme case takes place when the band wave functions are completely changed because of the core hole Coulomb potential (initial state approximation). Then the expression for the Green's function takes the form
\begin{equation}
\label{G_initial}
\tilde G^{-}_{\mu \nu }(E) = \sum_{n, \mathbf{k}} 
\frac{\langle \mu | \tilde \psi_{n \mathbf{k}} \rangle \,
\langle \tilde \psi_{n \mathbf{k}} | \nu \rangle}
{E-\tilde E_n(\mathbf{k})-E_{\rm f}-i \delta} \,,
\end{equation}
where $\tilde E_n(\mathbf{k})$ and $\tilde \psi_{\mathbf{k}}$ are perturbed band energies and wave functions, respectively. For PDOS in the presence of the core hole one gets
\begin{equation}
\tilde N_{\mu}(E) = \frac{1}{\pi} \, \mathrm{Im} \tilde G^{-}_{\mu \mu}(E)\, .
\label{dens2}
\end{equation}

MND theory~\cite{Mahan_67, Nozieres_69} considers the core hole filling completely in the dynamical way. The specific calculation scheme, which implements the MND theory in the framework of one-band model is presented in Ref.~\cite{Grebennikov_77a} In Refs.~\cite{Wessely_05, Wessely_06} a generalization of this scheme for the multiband x-ray absorption spectra calculations was performed. Following the notation of Refs.~\cite{Grebennikov_77a, Wessely_05} and neglecting the core hole natural width, we write the expression for the XEB intensity within the MND theory in the form 
\begin{eqnarray}
\label{I_MND}
I_{\rm MND}(E) \sim \mathrm{Re} \sum_{\mu, \nu} W_{\mu} 
\int \limits_{- \infty}^{E_{\rm f}} d E^{\prime} \, 
\tilde N_{\mu \nu}(E^{\prime})  
\int\limits_0^{+\infty} dt \, \times \nonumber \\
e^{-i \, (E+E_c-E^{\prime})\,t} \, \varphi_{\nu \mu }(E^{\prime},-t)\,,
\end{eqnarray}
where the matrix $\varphi(E,-t)$ satisfies the system of integral equations
\begin{eqnarray}
\varphi_{\mu \nu}(E,-t) = \delta_{\mu \nu} - 
\sum_{\mu^{\prime},\nu^{\prime}} 
V_{\mu \mu^{\prime}} \int\limits_{- \infty}^{E_{\rm f}}
dE^{\prime} \, \times \nonumber \\ 
K_{\mu^{\prime} \nu^{\prime}}(E,E^{\prime},t) \,
\varphi_{\nu^{\prime} \nu}(E^{\prime},-t)\,. 
\end{eqnarray}
The kernel of the integral operator $K_{\mu \nu}(E,E^{\prime},t)$ is given by
\begin{eqnarray}
K_{\mu \nu}(E,E^{\prime},t) = \sum_{\mu^{\prime}}
\frac{I_{\mu \mu^{\prime}}(E^{\prime},t)-e^{i (E^{\prime}-E) t} 
I_{\mu \mu^{\prime}}(E,t)}{E^{\prime}-E} \times \nonumber \\
\tilde N_{\mu^{\prime} \nu}(E^{\prime})\,,
\end{eqnarray}
where
\begin{eqnarray}
\label{I_fun}
I_{\mu \nu}(E,t) = \delta_{\mu \nu} - \sum_{\mu^{\prime}} V_{\mu \mu^{\prime}}
\int\limits_{-\infty}^{+\infty} d E^{\prime} \,
\frac{e^{i(E-E^{\prime})t}-1}{E-E^{\prime}} \, \times \nonumber \\
N_{\mu^{\prime} \nu}(E^{\prime}) \,,
\end{eqnarray}
\begin{equation}
\label{Core_potential}
V_{\mu \nu} = \sum_{n, \mathbf{k}} \left( \langle \mu | \psi_{n \mathbf{k}} 
\rangle 
\langle \psi_{n \mathbf{k}} | \nu \rangle E_n(\mathbf{k}) -  
\langle \mu | \tilde \psi_{n \mathbf{k}} \rangle \langle \tilde \psi_{n 
\mathbf{k}} 
| \nu \rangle \tilde E_n(\mathbf{k}) \right)\,.
\end{equation}
In the above expressions, the matrixes $N_{\mu \nu}$ and $\tilde N_{\mu \nu}$ have the form
\begin{equation}
N_{\mu \nu} = \frac{1}{\pi} \, \mathrm{Im} \, G^{-}_{\mu \nu }(E) \,, \qquad
\tilde N_{\mu \nu} = \frac{1}{\pi} \, \mathrm{Im} \, \tilde G^{-}_{\mu \nu }(E) \,.
\end{equation}
Note that the diagonal elements of matrixes $N_{\mu \nu}$ and $\tilde N_{\mu \nu}$ coincide with the PDOS defined above in Eqs.\,(\ref{dens1}) and (\ref{dens2}) in the final and initial state approximations.

There are two ways to define input data for the MND calculations expressed by Eqs.\,(\ref{I_MND})--(\ref{I_fun}). The first approach is to use the expressions (\ref{G_final}), (\ref{G_initial}), and (\ref{Core_potential}) to introduce the Green functions $G^{-}(E), \tilde G^{-}(E)$ and the core potential $V$. It works well if the band wave functions basis set is large enough. If it is not the case, the core hole potential $V$ depends on energy\cite{Wessely_06}. To minimize the error caused by the energy dependence of the potential one can use a second approach. The unperturbed Green function $G^-(E)$ and core hole potential $V$ defined from equations (\ref{G_initial}) and (\ref{Core_potential}), respectively, whereas the perturbed Green function $\tilde G^-(E)$ is obtained from the Slater-Koster equation\cite{Wessely_05}
\begin{equation}
\tilde G^-_{\mu \nu}(E) = \sum_{\nu'} G^-_{\mu \nu'}(E) \left[ T^{-1} \right]_{\nu' \nu},
\end{equation}
where
\begin{equation}
T_{\mu \nu} = \delta_{\mu \nu} + \sum_{\nu'} V_{\mu \nu'} G^-_{\nu' \nu}(E)
\end{equation}
Although the potential still depends on energy, the perturbed Green function now takes it into account.

\section{Auger-effect in the valence band}

The Auger process in the VB (radiationless decay of a hole state) can affect the K- and L-edge XEB of the metals. It leads to the suppression of the intensity at the bottom of the VB. As a result a tightened low-energy tail appears. Indeed, the initial state just before the emission process is characterized by a vacancy in the atomic core shell, which is filled by radiative transition of an VB electron. The resulting vacancy in the VB can be filled by Auger transition of an electron from the high-lying occupied VB states and the released energy is transferred to an Auger electron, which is excited into an empty state of the conduction band. The additional decay channel decreases the lifetime of the valence hole. Therefore the natural width of the VB vacancy is increased. Such mechanism of the VB hole states decay is especially important in metallic systems where there are no energy gaps.

Actually, this effect is implemented by the additional convolution of the spectrum $I(E)$ with the Lorentzian function $L(E', E)$, which full width at half maximum (FWHM) depends on energy. Following the described mechanism of VB hole decomposition we conclude that FWHM, $\gamma(E)$, is increased with the difference $(E_{\rm f}-E)$. Usually this dependence is approximated by a quadratic function~\cite{Livins_88}
\begin{equation}
\gamma(E) = \beta (E_{\rm f} - E)^2\,,
\label{gamma_E}
\end{equation}
where $\beta$ is a semi-empirical fitting parameter. In the framework of this approach the XEB including Auger correction $J_{\rm Auger}(E)$ is given by 
\begin{equation}
J_{\rm Auger}(E) = \int\limits_{-\infty}^{+\infty} dE^{\prime} \,
I(E^{\prime}) \, L(E^{\prime},E) \,,
\label{convolution}
\end{equation}
where
\begin{equation}
L(E^{\prime},E) = \frac{1}{\pi} \, 
\frac{\gamma(E^{\prime})/2}{(E^{\prime}-E)^2+ \gamma^2(E^{\prime})/4}\,.
\label{Lorentz}
\end{equation}
Fitting the $\beta$ parameter, which provides the best agreement between theory and experiment, the lifetime of the valence state $\tau(E)$ depending on energy might be obtained
\begin{equation}
\tau(E) \approx \frac{\hbar}{\gamma(E)}=\frac{\hbar}{\beta (E_{\rm f} -E)^2}\,. 
\label{tau_E}
\end{equation}
This expression is valid away from the Fermi level in order to avoid the infinity.

\section{Details of the calculations}

To account for the MND effect the DFT calculations were carried out using the projector augmented wave (PAW) method~\cite{Blochl_94} with plane wave basis set and the generalized gradient approximation parameterized by Perdew \textit{et al.}~\cite{Perdew_96} as implemented in the VASP program~\cite{Kresse_93, Kresse_96}. The supercells of a ($3\times 3\times 3$) periodicity, containing $54$ Mg atoms, and a ($4\times 4\times 4$) periodicity, containing $64$ Al atoms, were used. The cut-off energy was set to $400$\,eV in both cases, and the $k$-mesh sampling of the Brillouin zone was as dense as $9\times 9\times 15$ and $22\times 22\times 22$ for Mg and Al, respectively. In the XES simulations for graphene, the supercell was chosen to be as large as ($6\times 6$) and consists of the $72$ carbon atoms. The cut-off energy was set to $500$\,eV and the surface Brillouine zone was sampled with $30\times30$ $k$-points. The interplanar distance (vacuum region) was set to $15$\,\AA. 

To simulate the static electric field of the core hole the ``Z+1'' approximation was used for the central atom of the supercell. The perturbed local DOS $\tilde N_{\mu}$ was obtained for an atom in the center of the supercell and containing the hole. Unperturbed DOS $N_{\mu}$ was obtained using the same band calculation by projecting the Kohn-Sham band wave functions onto the valence atomic orbital of an atom situated far away from one with core hole. In order to get the appropriate input data for the MND calculations we used the second approach described in the end of Sec.\,\ref{par2}.

The expansion coefficients $c_{\mu n}(\mathbf{k})$ of the crystal orbital $\psi_{n \mathbf{k}}$ projected onto an atomic functions $\varphi_{\mu}$ defined by the expression
\begin{equation}
c_{\mu n}(\mathbf{k}) = \langle \varphi_{\mu} | \psi_{n \mathbf{k}} \rangle \,,
\end{equation}
were calculated using the approximate formula~\cite{Blochl_94}
\begin{equation}
c_{\mu n}(\mathbf{k}) \simeq  
\langle p_{\mu}^{\rm ps} | \psi_{n \mathbf{k}}^{\rm ps} \rangle \,.
\end{equation}
Here $\psi_{n \mathbf{k}}^{\rm ps}$ is the crystal pseudowave function, and $p_{\mu}^{\rm ps}$-- the so-called projection function of the PAW method~\cite{Blochl_94}.

The spectrum was convoluted with both a constant FWHM Lorentzian in order to take into account the natural width of the core level and with the energy depending FWHM Lorentzian to account for the Auger process in the valence band of metals. Furthermore, the spectrum was convoluted with a Gaussian to simulate the experimental instrumental broadening.

The values of natural widths of K and L levels were taken from Ref.~\cite{Krause_79}. The natural width of the metal L$_2$ and L$_3$ shells ($\le$~0.004 eV) appeared to be small in comparison with the instrumental functions FWHM and therefore it was not taken into account. The Gaussian FWHM was chosen equal to the original experimental spectra resolution in accordance with data published by other authors. All spectra are presented on a binding energy scale, with the Fermi level placed to zero.

\section{Details of the experiment}

L$_{2,3}$ XEB of the polycrystalline magnesium and aluminium were recorded at the modernized spectrometer RSL-1500 \cite{Lukirsky_70} with excitation of the ultrasoft X-ray spectra by an electron beam with an energy of $2.0$\,keV. As a detector a secondary electron multiplier SEM-6M with CsI photocathode in pulse mode of yield registration was used. The second detector with a beryllium bronze photocathode was situated on the zero photon beam (speculary reflected from the grating) and served as a monitor of the beam intensity. Spectra were recorded with a constant wave resolution and then, after accounting for the monitor intensity, were transformed to the constant energy resolution spectra.

The efficiency of the detector (the spectral dependence of the CsI quantum yield) was defined with respect to the quantum yield of gold which was considered to be a constant in the spectral range of $25-80$\,eV. The energy resolution of the obtained spectra was $0.04$ and $0.06$\,eV in the high-energy part of the spectra of magnesium and aluminium, respectively (up to $\approx\,2$\,eV from the edge of the L$_2$-XEB). The remaining (low-energy) part of the XEB of metals has a smooth shape almost independent of the energy resolution of the spectrometer if it does not exceed $0.4$\,eV. 

Obtained L$_{2,3}$ XEB had a very high contrast ratio (the ratio of the peak intensity of the bremsstrahlung intensity was more than $50$). After deducting of the bremsstrahlung background L$_{2,3}$ spectra were decomposed into L$_2$ and L$_3$ XEB components assuming the identity of their shapes.

\section{XEB of metals}

\begin{figure}
\centering
\includegraphics[width=0.55\textwidth]{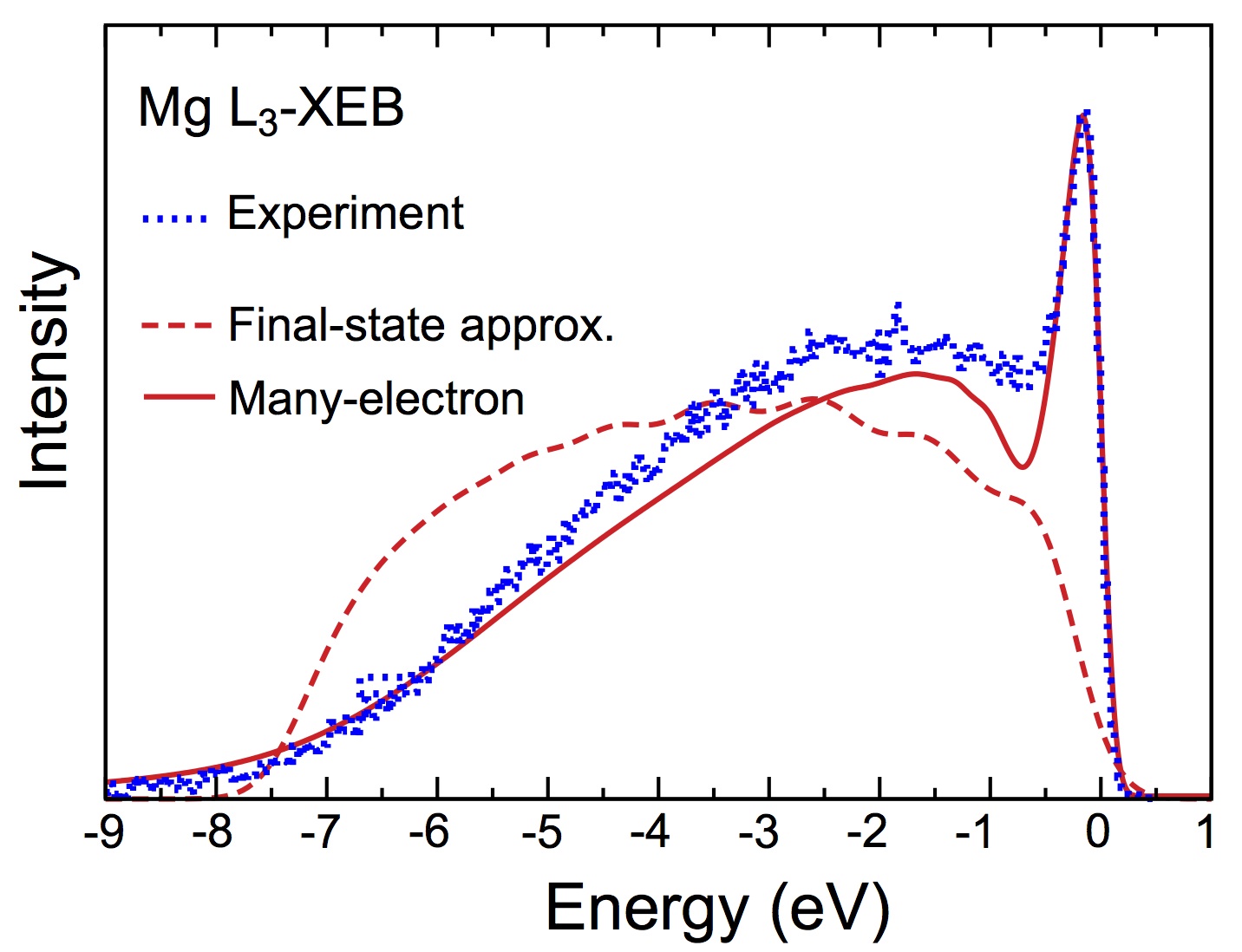}
\caption{L$_3$ XEB of magnesium metal: Experimental L$_3$ band (dots), theoretical spectrum in the final-state one-electron approximation (dash line) and when taking into account for the many-electron effects (solid line).}
\label{L_Mg_many}
\end{figure}

\begin{figure}
\centering
\includegraphics[width=0.55\textwidth]{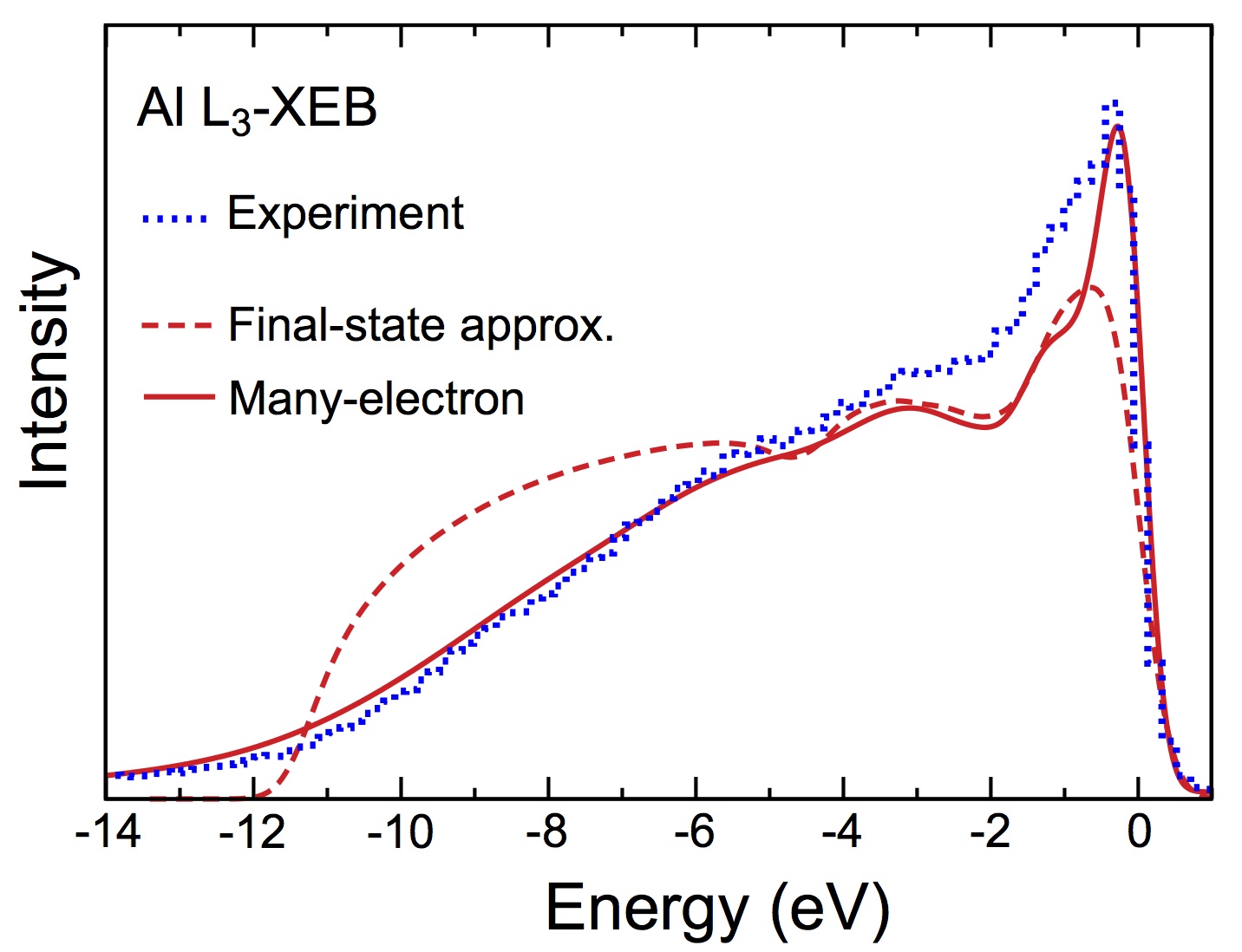}
\caption{L$_3$ XEB of aluminium metal: Experimental L$_3$ band (dots), theoretical spectrum in the final-state one-electron approximation (dash line) and when taking into account for the many-electron effects (solid line).}
\label{L_Al_many}
\end{figure}

In Figs.~\ref{L_Mg_many} and \ref{L_Al_many} the experimental L$_3$ XEB of magnesium and aluminium metals together with theoretical spectra calculated in the one-electron approximation and when accounting for the many-electron contribution (MND approach and Auger effect) are shown. 

One-electron spectra of Mg and Al were calculated by the Green's function method in the final-state approximation of the total static screening of the core vacancy and in fact they reflect the $s$-PDOS of VB of the corresponding metal. As was shown earlier~\cite{Ovcharenko_11}, the contribution of $d$-PDOS to L$_3$ XEB of Mg is small. Therefore, the single-electron emission spectrum of Mg obtained in the present study is in good agreement with the spectra calculated with dipole transition probabilities from $s$- and $d$-states of VB~\cite{Ovcharenko_11}.

As can be seen from Figs.~\ref{L_Mg_many} and \ref{L_Al_many} for L$_3$- XEB, the one-electron approximation poorly describes the features and the shape of the experimental spectra. Experimental curves in low energy parts fall off slower than theoretical spectra (in the one-electron approximation as well as in the MND approximation without Auger correction) and are distinguished by the presence of the tighten ``tails''. This behaviour may be explained by the Auger decay of vacancies in VB of the metals~\cite{Landsberg_49,Nemoshkalenko_74}. In addition, the experimental spectra have a narrow peak near the Fermi edge which is a result of a many-electron scattering of valence electrons on the core hole potential of metals (MND singularity~\cite{Nemoshkalenko_74, Nozieres_69}). The introduction of the many-electron corrections significantly improves the agreement with the experimental data. Many-electron theory well describes the behaviour of XEB  in the vicinity of the Fermi energy (peak intensity, its energy position, width) and in the low-energy part (shape and length of the tail). Note, that without Auger corrections the MND and one-electron spectra coincide with each other in the low energy region and differ at the Fermi level.

A good test of the validity of our calculations is the application of MND formalism for calculation of the K-emission bands. From the classical theory of MND  in the approximation of an isotropic electron gas~\cite{Mahan_67, Nozieres_69} follows that the MND singularity in K XEB should not occur. The intensity of K XEB near the Fermi level should be somewhat suppressed~\cite{Anderson_67} and gradually falls to zero. Indeed, it was shown that the screening valence band states have predominantly $s$-character. Since x-ray transitions from $s$-VB states onto $1s$ core-hole (K-XEB) are forbidden by dipole selection rules, dynamical screening is suppressed by orthogonality. Obviously, this is not the case for L-XEB (transitions onto $2p$ core-hole). In our approach the MND effect is taken into account in an \textit{ab~initio} way; therefore the degree of coincidence of the MND-calculated spectra with a single-electron and experimental XEB has to be high. According to Ref.~\cite{Krause_79} the natural width of the K levels of magnesium and aluminium are estimated to be $0.37$\,eV and $0.42$\,eV, respectively. In Ref.~\cite{Senemaud_71} the experimental Mg K-XEB was obtained with an energy resolution of about $0.6$\,eV, and Al K band resolution was $0.2$\,eV~\cite{Neddermeyer_73}. Therefore these values of the Gaussian FWHM were used for the convolution with 
theoretical spectra. 

\begin{figure}
\centering
\includegraphics[width=0.55\textwidth]{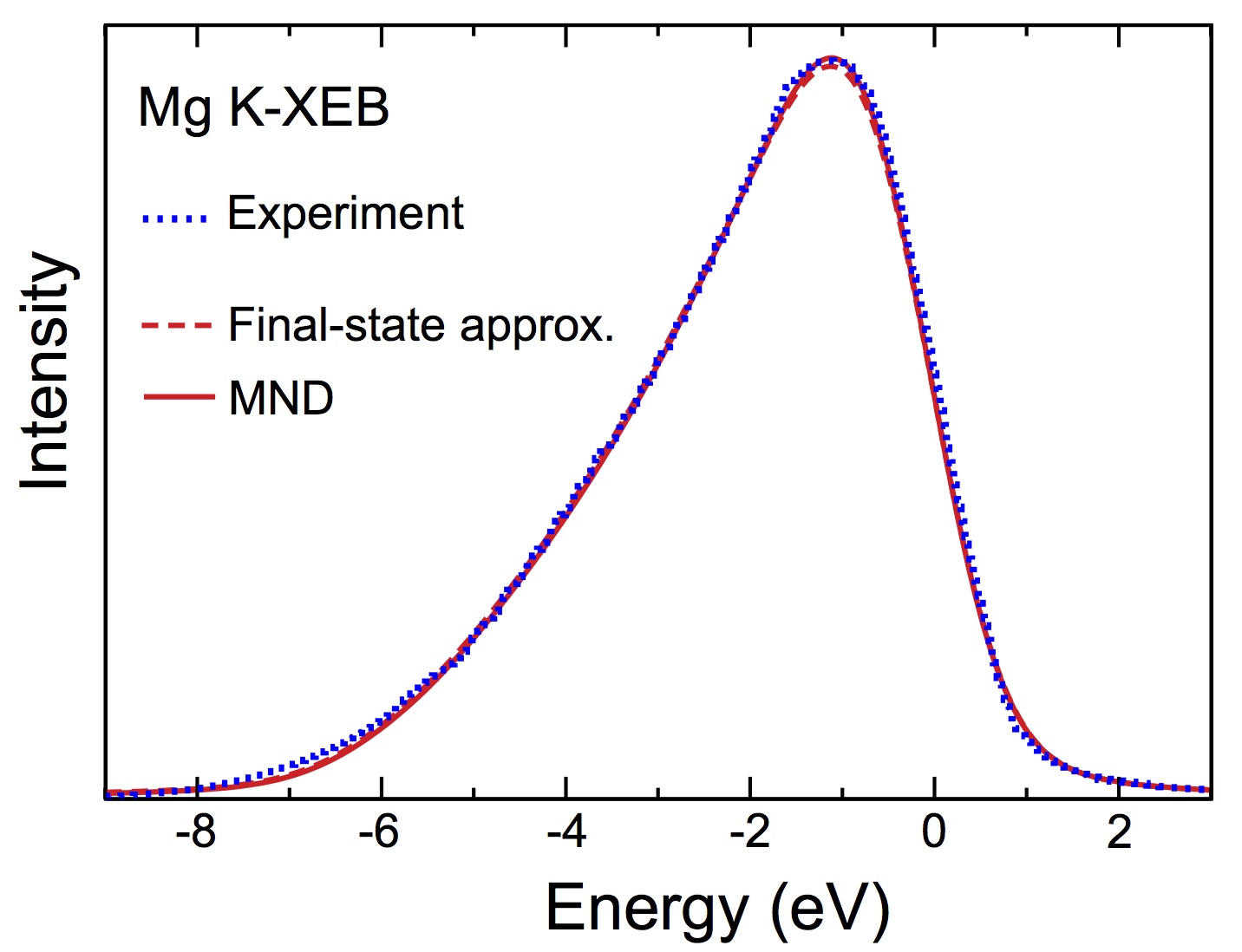}
\caption{K XEB of magnesium metal: Experimental K band (dots), theoretical spectrum in the final-state one-electron approximation (dash line) and when taking into account for the MND many-electron effects (solid line).}
\label{K_Mg_many}
\end{figure}

\begin{figure}
\centering
\includegraphics[width=0.55\textwidth]{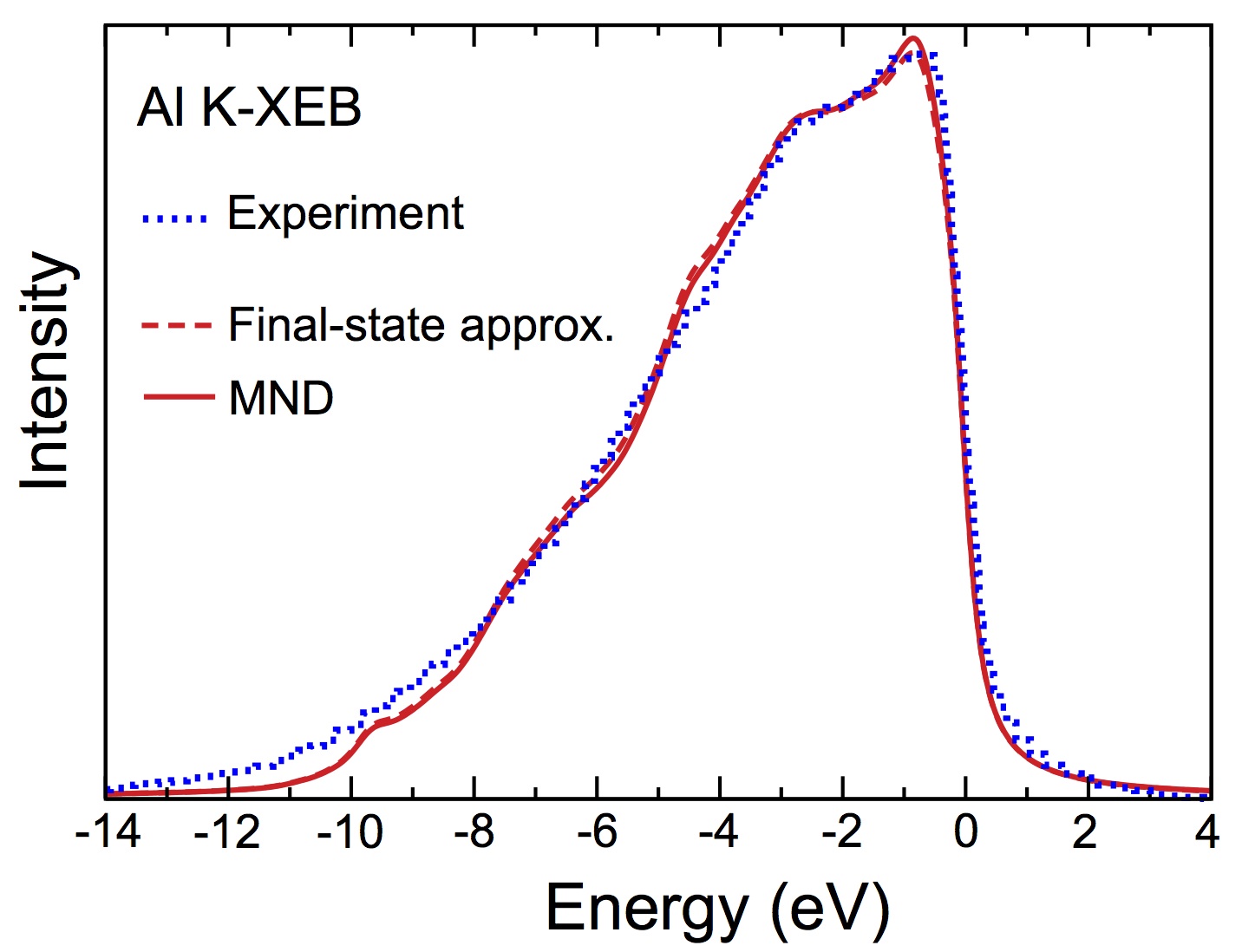}
\caption{K XEB of aluminium metal: Experimental K band (dots), theoretical spectrum in the final-state one-electron approximation (dash line) and when taking into account of the MND many-electron effects (solid line). }
\label{K_Al_many}
\end{figure}

As can be seen from Figs.~\ref{K_Mg_many} and \ref{K_Al_many} for K XEB of both metals, in contrast to L$_3$ XEB single-electron spectra and spectra calculated including the MND effect agree well with each other and with the experimental emission bands. Accounting of the distortions caused by the Auger effect in the VB is not required.

\section{The lifetime and natural width of levels in VB of metals}

In order to account for the influence of the Auger effect in the VB of metals on the shape of L$_3$ XEB, we calculated the convolution (\ref{convolution}) of MND XEB with a Lorentz function (\ref{Lorentz}) which takes into account the broadening of the valence levels due to existence of  the additional Auger decay channel. The functional dependence of FWHM on the level energy position was approximated by a quadratic function (\ref{gamma_E}) with a proportionality coefficient $\beta$, which was chosen for the best fit of calculated and experimental spectra. 

\begin{figure}[t]
\centering
\includegraphics[width=0.55\textwidth]{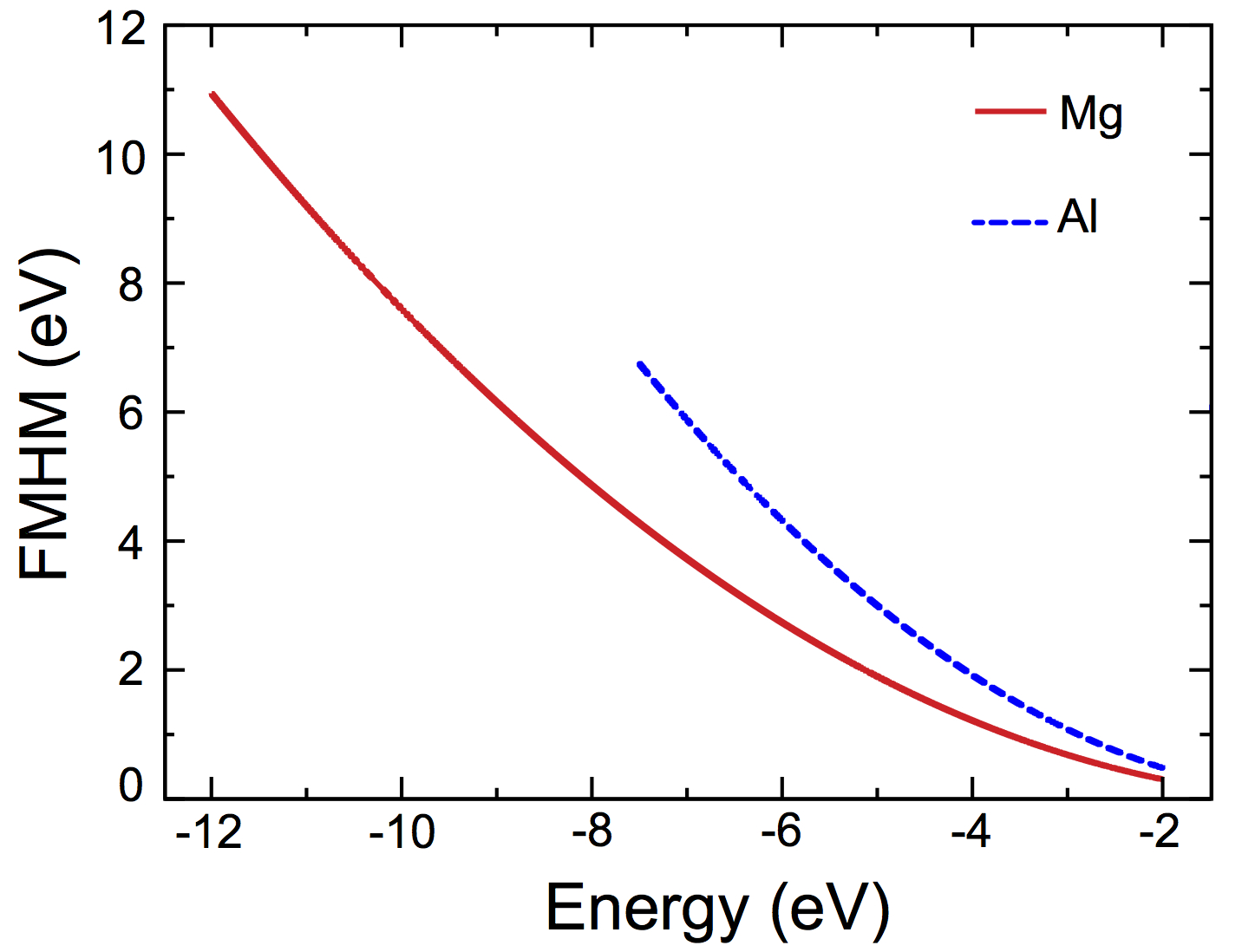}
\caption{The Lorentz width (FWHM) of levels in VB of metallic magnesium (dash line) and aluminium (solid line).}
\label{auger}
\end{figure}

For magnesium and aluminium the Auger effect was taken into account in the range of (E$_{\rm B}$,~E$_{\rm f}$-2eV), where E$_{\rm B}$ is energy of the bottom of the conduction band of the metal. This choice of the interval on the one hand allows to exclude from consideration the MND peak and on the other hand provides a smooth behaviour of the many-electron spectrum at the point of ``switch on'' of the Auger effect. The phenomenological parameter $\beta$ values for L$_3$- emission band of metallic Al and Mg are $0.038$\,eV$^{-1}$ and $0.060$\,eV$^{-1}$, respectively, and the widths of the valence levels as functions of energy are shown in Fig.~\ref{auger}. As can be seen from Fig.~\ref{auger} the probability of the Auger process in the VB of magnesium is larger compared to aluminium for the states with the same energy. However, due to significantly greater width of the VB and the L$_3$ XEB of aluminium, the Auger effect has a more pronounced influence on the shape of low-energy part of the Al L$_3$ XEB.

This estimation of the widths is correct only for $s$-states since only these states form the bottom of the crystal VB~\cite{Ovcharenko_11}. Indeed, contribution of $d$-states to the low energy parts of the VB and to the intensity of low energy part of Mg and Al L$_3$ XEB is negligibly low. As follows from Figs.~\ref{L_Mg_many} and \ref{L_Al_many}, Auger process significantly distort the shape of L-spectra of metals, but its influence on Mg and Al K XEB is very low if not negligible. This may be due to low density of $p$-states in the VB~\cite{Ovcharenko_11}. Calculated single-electron K-XEB themselves form  tight ``tails'' in the low-energy part of the spectra ($\sim E^{\frac{3}{2}}$), so the effect of the additional Auger broadening leads only to a slight delay in the ``tail'' while the low-energy region of L-spectra is deformed significantly. 

Such behaviour of $p$-PDOS and single-electron K-XEB in the low-energy part of the distributions may mask the distortions introduced by the realization of the Auger process in the VB. It is clear, however, if the Auger-process in the valence band is indeed possible it should take place for the initial states of any symmetry including the $p$-states. Therefore the absence of the requirement to introduce corrections for this many-electron effect into the K-XEB of magnesium and aluminium casts some doubt on the universality of the Auger mechanism of the decay of the final states of electron transitions from the valence band to the core shell vacancies of atoms in metals.

\section{Calculation of carbon K XEB in graphene}

As was already mentioned, according to the original theory of MND scattering in the free-electron approximation, the dynamical screening of the core hole by conduction electrons should not appreciably distort the K XEB and K absorption edge of metallic systems. However, as was shown in Refs.~\cite{Wessely_05,Wessely_06} the MND effect contributes significantly to the spectral distribution of the absorption coefficient of graphite near the C K-absorption threshold. On the other hand, the experimental carbon K-XEB of graphite~\cite{Muller_73} does not demonstrate any obvious signatures of the MND singularities. This asymmetry of MND scattering has been seemed us rather unexpected and it was decided to check its existence via calculations of the graphene emission and absorption spectra with insertion of the MND dynamical screening of the carbon $1s$-hole.

Graphite is a set of graphene layers which are shifted relative to each other forming the structure of the ABAB type. The interaction between layers is relatively weak and its existence possesses only small influence on the electronic structure of graphite which is almost completely determined by the electronic structure of one graphene layer. For this reason one can assume that the X-ray spectra of graphene and graphite should be very similar. K-edge absorption spectrum of graphene was calculated using the method described in Ref.~\cite{Wessely_05} and was almost identical to the K-edge absorption spectrum of graphite.

The shape of the experimental K-XEB of graphite, in contrast to X-ray spectra of Al and Mg, depends on the take-off angle of radiation (for emissions) and the angle of incidence (for absorption). In the experiments~\cite{Weser_10,Muller_73} these angles were $45^\circ$ and $63^\circ$ for the absorption and emission, respectively. Therefore instead of Eq.\,(\ref{probab1}) for the probability $W_{\mu}$ of intra-atomic transitions we used

\begin{equation}
W_{\mu} = \sum_{m_c} |\mathbf{e} \cdot \mathbf{A}_{\mu m_c}|^2\,,
\label{probab2}
\end{equation}
where the polarization $\mathbf{e}$ is perpendicular to $\mathbf{k}$, the direction of the X-ray photon propagation. For comparison with the experimental spectra of graphite, theoretical C K-spectra of graphene were convoluted with a Lorentzian function (FWHM=$0.2$\,eV, a lifetime width of the C $1s$ level in graphite~\cite{Sette_90}). Furthermore, a convolution with a Gaussian of $0.5$\,eV FWHM for the calculation of the emission band and $0.3$\,eV for the absorption spectrum have been made in order to account for the instrumental resolution.

\begin{figure}[h]
\centering
\includegraphics[width=0.55\textwidth]{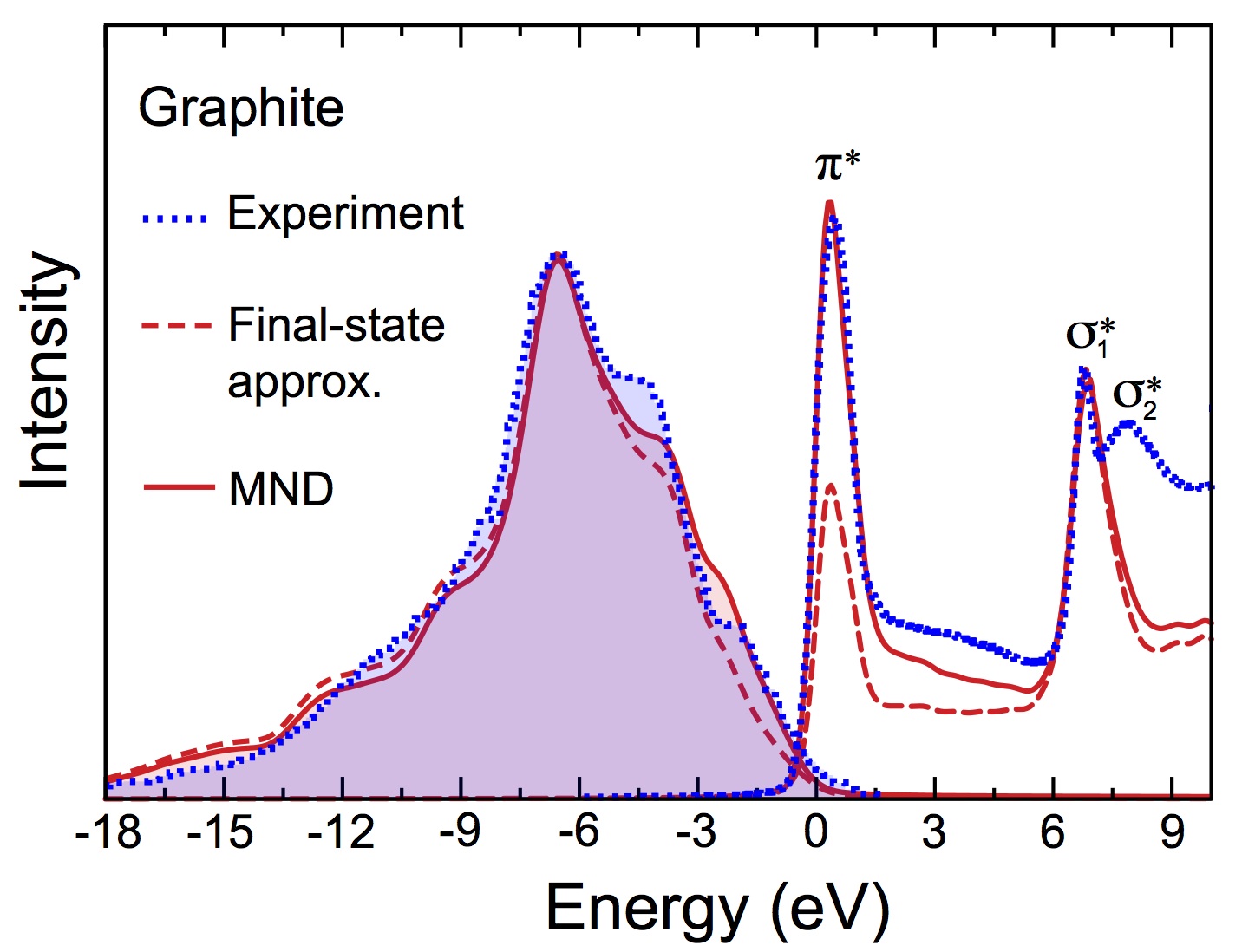}
\caption{Comparison of the experimental and calculated carbon K spectra of X-ray emission and absorption. Experimental K-absorption edge~\cite{Weser_10} and K- XEB~\cite{Muller_73} in the graphite are show by the dotted lines. Theoretical K-absorption spectrum and K-XEB in graphene in the final-state one-electron approximation are shown by dashed lines (Note: final-state approximations in the absorption spectrum is initial state approximation in the emission spectrum). Calculated K-absorption edge and K- XEB of graphene with the MND effect inclusion into consideration are shown by solid lines.}
\label{C_K}
\end{figure}

In Fig.~\ref{C_K} the K-band emission and absorption theoretical and experimental spectra of carbon in graphite are shown. It can be seen that the one-electron approximation satisfactory describes the experimental C K-XEB. Auger effect in the low-energy part of spectrum practically does not appear and does not require special consideration. The MND effect only slightly distorts the shape of the spectrum near the high-energy edge. This is not the case when analysing the absorption spectrum: one-electron approximation reproduces only the energy positions of $\pi^*$ and $\sigma_1^*$ peaks, but does not describe their relative intensities. 
Accounting for the MND scattering significantly improves agreement with experiment.

This so strong asymmetry in the manifestation of the MND singularities in the absorption and emission is observed for the first time in graphene. Obviously this effect is related to the peculiarities of the band structure of graphene near the Fermi level. Indeed, the linear electron dispersion in graphene leads to zero effective mass of electrons with energies close to $E_{\rm f}$. 

This means that the electronic subsystem instantaneously respond to any small changes in the crystal potential including a core shell vacancy creation caused by the absorption of X-rays. The suddenly appearing vacancy attracts a screening cloud of conduction electrons, which are excited from the occupied states below the Fermi energy into the conduction band states above it. This local increase in the unoccupied DOS near $E_{\rm f}$, in which the core electrons can be optically excited, leads to appearance of the MND singularity in the absorption spectrum of graphene. In other words, the near edge X-ray photon absorption takes place in a dynamical screening of carbon $1s$ vacancy conditions. In the emission spectrum a core hole (the initial state of the radiative decay) is almost completely screened due to the high mobility of the conduction electrons near the Dirac point. Therefore the radiative X-ray transition arises in the conditions of the almost complete static screening of the $1s$-vacancy and does not influenced by the dynamical MND scattering.

This effect may be important in characterization of the conduction electrons dynamics, their mobility and the characteristic relaxation times in graphene-like systems.

It is clear from Fig.~\ref{C_K} that neither one-electron nor MND spectra do not describe experimental $\sigma_2^*$ peak of the K absorption of graphene. The detailed discussion of this phenomenon was carried out in Ref.~\cite{Wessely_05} Briefly, this comes from the fact that the $\sigma_2^*$ peak is due to nonexcitonic states, which are delocalized. Therefore it cannot be described neither by MND theory (which take into account only exitonic states) nor by final-state one-electron approximation (in the static core-hole field). On the other hand, this peak can be taken into account by initial-state one-electron approximation~\cite{Wessely_05,Wessely_06}.

\section{Conclusions}

In order to account for the influence of the many-electron effects of the core hole dynamical screening by valence electrons on the shape of characteristic XEB of metals we combined MND theory with the \textit{ab~initio} electronic structure calculations of crystals. Auger effect in the valence band was taken into account in a phenomenological way. Within these approaches we have calculated K and L$_{2,3}$ XEB of Mg and Al crystals. Extremely good agreement between theoretical and experimental spectra was found across the whole spectral range of the bands. Process of the core vacancy dynamical screening significantly distort the shape of L$_{2,3}$ XEB of metals in the vicinity of the Fermi energy, while the Auger processes in the valence band give rise to tighten low-energy ``tail'' at the bottom of the bands. In the framework of the Auger effect model used, the energy dependences of the $s$-hole states natural width in the valence bands of magnesium and aluminium were calculated.

The developed approach was applied for the calculation of the carbon K-edge emission band and absorption spectrum of graphite. It was found that experimental C K-band is close in shape to the single-electron XEB. Accounting for dynamical screening of C $1s$ vacancy brings only minor changes to the high-energy part of XEB. In contrast, MND scattering induces a very prominent contribution in the K-absorption spectrum near the absorption edge. The observed asymmetry of C $1s$ hole dynamical screening in the absorption and emission processes can be related to the high mobility of VB electrons near the Dirac point of graphene. This interesting effect will be investigated in more detail in future publications.

\section*{Acknowledgements}

This work was supported by St. Petersburg State University grants No 11.37.24.2011, RFBR grant 
No 12-03-01140-a and G-RISC-DAAD project No C-2011b-3.

\end{document}